\documentstyle[12pt, aaspp4]{article}
\begin{document}
 
\def\today{\number\year\space \ifcase\month\or  January\or February\or
        March\or April\or May\or June\or July\or August\or
September\or
        October\or November\or December\fi\space \number\day}
\def\fraction#1/#2{\leavevmode\kern.1em
 \raise.5ex\hbox{\the\scriptfont0 #1}\kern-.1em
 /\kern-.15em\lower.25ex\hbox{\the\scriptfont0 #2}}
\def\spose#1{\hbox to 0pt{#1\hss}}
\def\simlt{\mathrel{\spose{\lower 3pt\hbox{$\mathchar''218$}}
     \raise 2.0pt\hbox{$\mathchar''13C$}}}
\def\simgt{\mathrel{\spose{\lower 3pt\hbox{$\mathchar''218$}}
     \raise 2.0pt\hbox{$\mathchar''13E$}}}
\def\etal{et al. }

\title{HD 209458: Physical Parameters of the Parent Star and the
Transiting Planet}
\author{Ann Marie Cody \& Dimitar D. Sasselov\altaffilmark{1}}
\affil{Dept. of Astronomy, Harvard University, 60 Garden St., Cambridge MA 02138}
\altaffiltext{1}{Alfred P. Sloan Foundation Fellow}

\begin{abstract} The Sun-like star HD~209458 harbors a close-in giant
planet that transits across the star's disk, and thus allows an
unprecedented access to the basic parameters of the planet, given a
certain knowledge of the basic parameters of the star, namely its mass and
radius. We present theoretical stellar evolution model calculations for
HD~209458 and discuss the uncertainties involved in deriving the stellar
mass and radius. We derive the mass, $M=1.06~M_{\odot}$, radius,
$R=1.18~R_{\odot}$, and age, $t=5.2$~Gyr of the star with uncertainties of
10\% or more. The dominant sources of uncertainty remain to be the helium
abundance estimate and the treatment of convection, even after an
optimistic estimate for the effective temperature of the star. However, we
find that in deriving the radius of the planet, $R_p$, the relevant
stellar model input is the M/R relation, which runs orthogonal to a
degeneracy in the transit light curve solution and greatly improves the
estimate of $R_p$. Theoretically the M/R relation has a lower uncertainty
than the $M$ and $R$ separately. We estimate the planet radius and mass to
be $R_p =1.42^{+0.10}_{-0.13}~R_{J}$ and $M_p =0.69\pm0.02~M_{J}$.

\end{abstract}

\section{Introduction} The transit detection of HD~209458~b
(Charbonneau et al. 2000; Henry et al. 2000) was a milestone in the study
of extrasolar planets. Since the first close-in extrasolar giant planet
(CEGP), 51 Peg b, was discovered in 1995 by Mayor \& Queloz (1995), the
nature of these unusual type of objects was not truly understood. Eleven
more close-in extrasolar giant planets with orbits $\leq 0.05$~AU are now
known. However, thanks to the transit detection of HD~209458~b, we are certain
now that they are indeed gas giant planets, much like our own Jupiter and
Saturn. The transit of HD~209458~b fixes the orbital inclination (which
removes the $\sin~i$ ambiguity in mass), it gives the planet radius, provides
the average planet density, and confirms that the CEGPs are gas giants.

The crucial parameters for understanding the nature of the extrasolar
planet HD~209458~b are its mass and radius, yet an accurate determination 
depends on
our independent knowledge of the mass and radius of its parent star HD~209458
(Charbonneau et al. 2000; Mazeh et al. 2000). The latter is a solar-like
star --- a G0 subgiant (slightly more massive and evolved than our
Sun),
and our means to determine its basic physical parameters should be fairly good.  
However, the demands on precision are so high that current stellar
evolution codes and model atmosphere analysis can hardly live up to the
challenge. In this paper we make a detailed analysis of the stellar interior
evolution models for HD~209458, relying on the $Hipparcos$ parallax for
its luminosity, $L$, and on the spectroscopic analysis of Mazeh et al. (2000) 
for its effective temperature, $T_{\rm eff}$, and metallicity, $Z$.
Our main goal is to quantify the uncertainties and systematic errors that
are involved.

An alternative approach is possible: either with multi-color transit photometry
(Jha et al. 2000; Deeg et a. 2001), or with very precise transit photometry
(Brown et al. 2001). These studies assume that the stellar mass, $M$, is known
and derive $R$, $R_p$, $i$, and $u$ (a one-parameter description of limb
darkening). The procedure reduces the apparent uncertainty in $R_p$,
but since $R$ and $i$ remain strongly correlated, we suggest an improved
approach. Our approach relies on the astrophysical correlation
between stellar $M$ and $R$, which breaks the degeneracy. In addition,
it is a more robust result of stellar evolution theory than the mass estimate.

\section{The Model} 
The star HD~209458 is extremely similar to our Sun from the point of
interior modeling (Figure 1). Such close similarity allows us to use
a theoretical model that is very well constrained for the Sun. For many
parameters the changes to the HD~209458 interior model are small and
we can use solar constraints to test their reliability, e.g., the
size of the convection zone, etc.
Our code is largely based on the Sienkiewicz, Paczynski, Ratcliff (1988) 
code as updated and distributed by R. Sienkiewicz, where our
changes concern mainly the upper boundary and convection. It is a
Henyey code that solves the equations of stellar structure in one
dimension. We use the OPAL equation of state (Rogers et al. 1996) and
the latest Livermore opacity tables (OPAL96, Iglesias \& Rogers, 1996) for 
the Grevesse \& Noels (1993) heavy element mixture. The tables are augmented
by the Alexander \& Ferguson (1994) data on molecular and grain
opacities. The nuclear reaction rates are calculated according to Bahcall 
\& Pinsonneault (1995). The diffusion of hydrogen and oxygen is
treated after Thoul, Bahcall, \& Loeb (1994). The upper boundary is
an atmosphere with a temperature distribution from model atmosphere
integration and Kurucz (1992) opacities; for models very different from the
present Sun, the Eddington $T-\tau$ relation is used. Convection is
treated with the standard mixing-length prescription and the
Schwarzschild stability criterion. Slow rigid-body rotation is 
allowed. The evolution of the abundances of H,$^3$He,$^4$He,$^{14}$N,
$^{16}$O, and $^{17}$O is followed. Magnetic fields are not considered.

The ZAMS models are computed with initial guessed values of the boundary
parameters taken from the standard solar model for $M_*$=1.0~$M_{\odot}$
and scaled for masses in the Sun's vicinity. The subsequent iteration is
similar to that used by Ford, Rasio, \& Sills (1999).

As a check we compared runs with the Yale Rotating Evolution
Code (YREC) for as close as possible initial parameters and overall
conditions prepared by D. Guenther (2000, private communication).
Despite many differences in assumptions and numerics, the
codes compare very well and we use the runs to study possible
systematics in our theoretical HD~209458 model (see \S 3.3).

\section{Stellar Parameters}
\subsection{Observational Data}

$\quad$ While the evolution code is capable of outputting very precise
values for the stellar properties in question, the accuracy of our results
was contingent upon the quality of observational data.  An examination of
the literature revealed very few studies that have provided physical
properties for HD 209458.  Fortunately, however, our work followed on the
heels of an extensive spectral analysis performed by Mazeh \etal (2000).  
They derive an effective temperature of $6000\pm50$~K and a metallicity
([Fe/H]) of $0.00\pm0.02$.  We adopted these values but explored a wider
range of heavy element abundances than the [Fe/H] measurement error
quoted above. 

  Another crucial parameter under consideration was the stellar
luminosity.  We have put to use recent data provided by the $Hipparcos$
Space Astrometry Mission (ESA 1997, SP-1200), which measured an apparent V
magnitude of 7.65 and a parallax of $21.14\pm1.00$ mas for HD~209458.  
Assuming negligible interstellar extinction, this implies an absolute
magnitude of $4.28\pm0.10$ in V.  To transfer from magnitude to luminosity,
we applied a bolometric correction (BC), relying on the $T_{\rm eff}$/BC
scales derived by Flower (1996).  The star's derived temperature of
6000~K requires a correction factor of $-0.45\pm0.007$. These
values indicate a bolometric magnitude of 4.23, and hence a luminosity
$1.61\pm0.15 L_{\odot}$ (or $0.208\pm0.040$ in logarithmic units).

\subsection{Theoretical Evolution Results} $\quad$ With all of the
requisite physical parameters in hand, we began the stellar
modeling process. To guide interpretation of the program output, we
constructed a portion of the H-R diagram with a central
temperature/luminosity error box given by our adopted uncertainties
(see figure 1).  The limits of this box, 3.7745 to 3.7818 for
temperature (log units), and 0.168 to 0.248 for luminosity (log
units), indicate the narrow range of "target" values imposed by
observation. In the search for plausible evolutionary tracks, we
separated the code routine into runs of specific metallicities,
including $Z$~=~0.013, 0.015, 0.016, 0.020, 0.025, and 0.031. These
abundances{\footnote{Where $\log Z=\log\frac{X}{X_{\odot}} + \log Z_{\odot} +
[Fe/H],$ and $X_{\odot}$, $Z_{\odot}$ are the solar abundances of hydrogen 
and metals (by mass) for $X+Y+Z=1$.}}
were chosen based on the adopted metallicity range
[Fe/H]~=~$0.00\pm0.1$. We elected to
use the precise solar values $X_{\odot}=0.7059$ and $Z_{\odot}=0.0200$
(Guenther \& Demarque 1997), and we assumed the stellar initial
hydrogen content $X=0.7$ for
HD~209458. The parameters $Z$~=~0.016 and $Z$~=~0.025 were taken to
be the lower and upper bounds of allowable metallicities, but a few
models with the values 0.013 and 0.031 were used to test more
extreme evolutionary behavior.

  After computing a grid of models based on the chosen metal
abundances, we determined the most favorable evolutionary tracks by
retaining only those that met the strict temperature and luminosity
requirements. The results of our modeling can be seen in figure 1; a
number of theoretical tracks pass through the error box.  In
proceeding to run the evolutionary code on a wide variety of stellar
parameters, we succeeded in associating with each metallicity a range
of masses that yielded acceptable temperatures and luminosities. We
have plotted tracks for the metallicities 0.016, 0.020, and 0.025 in
figure 1. Masses shown for $Z$~=~0.016 are (in order of decreasing
temperature) 1.009, 0.990, and 0.978$~M_{\odot}$.  The metallicity of
0.02, constituting our best estimate for HD~209458 as well as the
canonical solar value, was implemented for models of mass 1.15, 1.10,
1.09, 1.06, 1.05~$M_{\odot}$ (the track for 1.10 is omitted from
figure 1). It was also used in constructing a zero-age main sequence
(ZAMS) track that can be seen cutting across the left portion of
figure 1.  The upper range of metallicities, 0.025, was run through
the code with masses of 1.175, 1.13, 1.12, and 1.10~$M_{\odot}$ (1.12
is not plotted in figure 1). A more complete set of data from the
evolutionary tracks is available in table 1.  As evident from the H-R
diagram, all of the metallicities we have selected have some range of
masses that produces models reaching the desired temperature and
luminosity (i.e. passing through the center of the error box).  While
not obvious from the figure, HD~209458 must have a metallicity as high
as $Z$~=~0.05 (assuming solar hydrogen abundance) before its
evolutionary models no longer intersect the error box.  This situation
allows for great freedom in choosing combinations of masses and
abundances that match observational data.  Therefore, narrowing down
the parameters further required restrictions on mass and age.

 The age of HD~209458 is not known; we could only put rough constraints
on it from observed levels of stellar activity. For HD~209458 such
constraints happen to make little difference, being somewhere
between 4 and 7 Gyr of age (Mazeh \etal 2000). 
However, it is instructive to see how
derived ages correlate with the possible range of stellar metallicity,
and eliminate some models that are extraordinarily old or young.
For a particular $Z$-value, we
located three evolutionary tracks of differing mass: one passing
directly through the target temperature of 6000~K and luminosity
$1.61~L_{\odot}$, one that simultaneously reached the highest
allowable temperature and lowest luminosity (thus giving the
lowest age), and one that reached the highest luminosity and
lowest temperature (giving the greatest age).  We then noted the
ages for which each model touched the limits of the error box.  
In the interest of making visual comparisons, we have plotted all
metallicities versus these sets of age extremes in figure 2.

  The results of our modeling reveal that not all of the theoretical
evolutionary tracks for HD~209458 that pass through the
temperature/luminosity error box evolve to a desirable age.  On the
scales that we are interested in, age range and $Z$ follow an
approximately linear relationship.  In fact, no models with metallicity
greater than 0.030 ([Fe/H]~=~+0.18) or less than 0.012 ([Fe/H]~=~-0.22)
can be considered viable.  Therefore, we can be confident that
HD~209458's metallicity is within these two values.  Given the star's
predicted solar metal abundance, this is hardly surprising; the
theoretical models achieving optimal temperatures and luminosities are
within a credible zone of abundances.

  Both the evolutionary tracks and the analysis of stellar age range
lend support to the original temperature and luminosity data for
HD~209458: 6000K (log~=~3.778) and 1.61~$L/L_{\odot}$ (log~=~0.208).  
These are the values at the center of our error box, and a model with
$Z$~=~0.02 reaches them when computed with a mass of 1.06~$M_{\odot}$.  
It should also be noted that models of mass 1.09 and 1.04~$M_{\odot}$
just barely reach the range of reasonable temperatures and luminosities.  
All three models remain comfortably within the 4-7 Gyr range.  The
parameters M~=~$1.06^{+0.03}_{-0.02}~M_{\odot}$, $\log T_{\rm
eff}~=~3.778\pm0.004$, and $\log L/L_{\odot}~=~0.208\pm0.040$ then
describe our best stellar model for HD~209458 with solar metallicity.  
If we allow for our assumed range of
metallicities, $Z=0.02\pm0.005$ ($[Fe/H]=0.0\pm0.1$), then the mass
uncertainty increases to $^{+0.11}_{-0.09}$. This is identical to
the mass uncertainty ($\pm 0.10$) estimated by Mazeh \etal (2000).
Propagating all of our
errors, we find a radius of $1.18^{+0.07}_{-0.08}~R_{\odot}$.
The age associated with this combination is 5.2 Gyr.
Our next step is to take account of other systematic uncertainties.

\subsection{Uncertainties}

There are three large sources of uncertainty. They have to do with
the stellar helium abundance (the surface amount of helium cannot
be observed in cool stars), the diffusion of helium and other heavy
elements, and the treatment of convection.

{\em Helium.} The initial abundance of helium, $Y$, is an important yet
unobservable parameter that influences the stellar evolution
model.  Thus, we examined the effect of shifting the degree of
our models' helium enhancements.  Figure 3 illustrates the change
induced in temperature when $Y$ is varied by 0.02 (from the
original figure of 0.28), and metallicity is held at a constant
$Z$~=~0.02.  For a helium abundance of 0.26, overall model
temperature decreases, and the mass must be raised by
$~0.04~M_{\odot}$ in order for the evolutionary track to once
again pass through the center of the temperature/luminosity error
box.  For a higher helium abundance of 0.30, the opposite is true;
temperature increases, and the model mass must be decreased by
$~0.04~M_{\odot}$ to achieve the optimal temperature and
luminosity.  In addition, an altered $Y$-value will cause a slight
shift in the stellar age.  Our models indicate that an increased
helium abundance corresponds to an increase in age for models of
similar temperature and luminosity.  The best-fit model with
$T_{\rm eff}$~=~6000~K and L~=~1.61~$L_{\odot}$ increases to 5.7
Gyr for a helium abundance of 0.30, and decreases to 4.7 Gyr for
an abundance of 0.26. In summary, $Y$ contributes about 4\% as a 
systematic uncertainty to the derived stellar mass for HD~209458.

{\em Convection.} In using a local prescription to compute the
stellar convective envelope $-$ the mixing length theory, we have
to specify the free mixing length parameter, $\alpha$.
Since $\alpha$ is a number
unavailable observationally, we are left to surmise that HD~209458
is similar to the sun in its envelope convection properties.  
The comparable temperatures and metallicities of the two stars
makes it likely that they share an $\alpha$ of 1.69 (which is the
standard solar value (Guenther \& Demarque 1997).  
Nevertheless, it is instructive to investigate several
evolutionary models with several different $\alpha$-values, following
the findings of multi-dimensional radiation hydrodynamics models
of stellar convection (e.g., Ludwig, Freytag, \& Steffen 1999).  The
evolutionary tracks plotted in figure 4 correspond to mixing
length parameters of 1.40, 1.55, 1.69 (solar), 1.85, and 2.0, and
they reveal that variations in $\alpha$ do indeed have a
noticeable effect.  Increasing the value by 0.15 leads to an
overall temperature increase of about 75~K. The mass must be
augmented by 0.03~$M_{\odot}$ to force the model back through the
center of the $T_{\rm eff}$/L error box.  For smaller $\alpha$'s,
the reverse trend occurs.  Even more significant, however, are the
shifts in age.  For a set of models passing through the center of
the box, decreasing the mixing length parameter to 1.40 pulls
HD~209458's age down to 1.8 Gyr, and increasing $\alpha$ to 2.00
inflates it to 8.2 Gyr. Both numbers are clearly extreme values.
The more moderate $\alpha$~=~1.55 and
$\alpha$~=~1.85, on the other hand, produce modest jumps in age of
about 1.5 Gyr.  Although it is unlikely that extreme mixing length
parameter values are the case, they are certainly incompatible
with our established temperature and luminosity, unless [Fe/H] is
far different from solar. Therefore the unknown $\alpha$ contributes
at the 3\% level to the uncertainty of the stellar mass (and much
more to the stellar age).

{\em Diffusion.}  Another systematic uncertainty we can quantify
is the magnitude of heavy element diffusion. Not accounting for the
diffusion of helium produces the offset ($\sim$0.2\%) shown in figure 1 between
the known temperature and luminosity of the Sun and its theoretical
evolution track. We ran the code for the
standard solar values of $X$~=~0.7059, $Z$~=~0.02 (Guenther \&
Demarque 1997), and $\log T_{\rm eff}~=~3.7612$.
A run with YREC reproduced precisely the effect as well.
In summary, the systematic uncertainty due to diffusion is much less
substantial (below 1\%), than the two discussed above.

To combine all uncertainties involved, we should note the following.
The region of the H-R diagram inhabited by HD 209458 is populated by
virtually vertical evolution tracks for our range of $Z$ and $Y$ (Figure 1).
Therefore the stellar mass determination is subjected to the largest
uncertainty, which could only be remedied by a better $T_{\rm eff}$ $or$ age
derivation. Unfortunately, a better parallax would not help. A more
accurate $T_{\rm eff}$ cannot be derived spectroscopically at this time;
even the 50~K uncertainty used here is quite optimistic.
The systematic uncertainties due to $Y$ and $\alpha$ also affect
primarily the stellar mass determination. Therefore we have finally:
$M=1.06\pm0.10(obs)\pm0.07(sys)~M_{\odot}$ for HD 209458's mass.
On the other hand, the stellar radius remains: $R=1.18\pm0.10~R_{\odot}$, and
could be improved by a better distance to HD 209458, $e.g.,$ a FAME
parallax. Clearly the linear correlation between $M$ and $R$ (i.e., the
ratio M/R) in that
small region of parameter space is the most accurate outcome of the
theoretical models, because the model systematics are minimized.

\subsection{Planetary parameters} $\quad$ To derive the planetary radius
($R_{p}$) and orbital inclination ($i$) from the stellar radius and mass,
we employed the light-curve data of Brown \etal (2001).  Following
the method discussed in Sackett (1999) and used by Charbonneau \etal
(2000) and Mazeh \etal (2000), we constructed the appropriate
limb-darkened transit models for HD~209458. For this particular analysis,
we adopted the same limb darkening parameter, $c_{\lambda}$, as Mazeh
\etal (2000): 0.56. In addition, we used an orbital period of 3.524739,
obtained by Robichon \& Arenou (2000) through $Hipparcos$ data.  With
these values, we produced a series of transit light curves for selected
planetary radii and orbital inclinations and calculated the $\chi^{2}$
statistic for each, based on the Brown \etal (2001) photometry.  
For $R$~=~1.18 and $M$~=~1.06, we derive a planetary radius of $R_{p}=
1.42^{+0.10}_{-0.13}~R_{J}$, very close to the results of Mazeh \etal (2000)
and Brown \etal (2001).  The inclination angle, $i$, is
$86.1^{+1.1}_{-0.5}$.

We can decrease the uncertainties in $R_{p}$ and $i$ by a factor of 2 for
HD~209458 by using the theoretical M/R ratio for the star. 
Two reasons are responsible.
First, as we noted above, the relation minimizes two sources of systematic
errors on the stellar mass. Second, and more importantly, the theoretical
M/R relation in this small region of the H-R diagram breaks a degeneracy in
$R$, $i$, and $R_{p}/R$. The degeneracy is well-known in
light curve solutions of detached eclipsing binary stars with a transit
eclipse of a limb-darkened primary (e.g., Popper 1984). The times of second
and third contact are poorly determined, unless the limb darkening is precisely
known (or zero). Often the mass of the primary star, $M$, is used to derive
the orbital velocity of the secondary (unknown for our planet) in the
calculation of the latitude of the transit. The transit latitude relates
$M$, $R$, and $i$, and the best fit for $R_{p}/R$ is aligned along a
positively correlated $M$ \& $R$. This positive M/R correlation can be
seen in the HD~209458 transit curve solutions of $HST$ photometry
(Brown \etal 2001; their Fig. 5) and multi-color photometry (Deeg \etal
2001; their Fig. 5). The latter is $R=0.34M+0.825(\pm 0.06)$. However, in the
vicinity of HD~209458, the theoretical relation we compute has a negative
slope, thus nicely constraining the transit curve
solution, as shown in Figure 5.
In the neighbourhood of the best value, we have:
$$
\frac{M}{M_{\odot}}-1.06 = -0.96\biggl(\frac{R}{R_{\odot}}-1.18 \biggr)
 -2.05\biggl(Y-0.28 \biggr) +15.39\biggl(Z-0.02 \biggr).
$$
The correlation between stellar mass and radius for normal single stars
near the main sequence (masses 1.0 to 1.1$M_{\odot}$) at {\em constant}
luminosity would emerge from the definition of effective temperature.
A hot main sequence star would be smaller than a cool main sequence star
of the same $L$, $Y$, and $Z$ as shown in Figure 1. 
Of course, their ages will differ. Note that the observational determination of
$R_{p}$, from the analysis of the transit light curve, is done by holding 
$L$, $Y$, and $Z$ fixed. With virtually vertical evolution tracks,
the stellar mass, $M$, used in the solution, is practically independent of
the stellar $L$ in the error box of HD~209458. Incidentally, $Z$ is mostly
correlated to the $T_{\rm eff}$ in its derivation, as well.

\section{Discussion \& Conclusion} $\quad$ By exploring a wide variety of
physical parameters characterizing HD 209458,  we have been able to
confirm the general validity of previous metallicity measurements as well
as stellar mass and radius calculations, noting the consistencies in age
within the regime of $T_{\rm eff}~=~6000K$ and $L/L_{\odot}~=~1.61$.  
Although the metallicity we have adopted may require adjustments, the
models are enough to rule out the possibility that it is far different
from solar.  With a full set of evolutionary tracks in hand, we can hold
the results of $M=1.06\pm0.13M_{\odot}$ and $R=1.18\pm0.10R_{\odot}$
to a high level of confidence.
For comparison, the existing theoretical mass and radius determinations
for HD~209458 make use of precalculated evolution model isochrones.
Mazeh \etal (2000) used four different sets of models $-$ by Schaller \etal
(1992), Bertelli \etal (1994), Claret (1995), and by Yi, Demarque, \&
Oemler (1997). And in a survey of several hundred thousand stars Allende Prieto
\& Lambert (1999) derived a mass and a radius for HD~209458 using isochrones
from Bertelli \etal (1994). Our results differ only in the acknowledgment
of two sources of systematic error (unknown $Y$ and $\alpha$) which increase
the mass and radius uncertainty for HD~209458.

  The availability of new, precise transit photometry (Brown \etal 2001)
only reinforces further
the need for better stellar structure models. Our analysis
shows that the relevant model input for solving transit light curves
is the relation between stellar mass and radius in \S3.4. For future
transit systems, such mass-radius relations should be derived uniquely from
the corresponding evolution models for the parent star.
From our derived inclination angle $i$ of $86.1^{o}$, we
obtain a planetary mass (dependent on our estimated stellar mass,
$1.06M_{\odot}$) of $0.685\pm0.02M_{J}$. With the derived radius,
$1.42^{+0.10}_{-0.13}R_{J}$, the planet density would be only
$0.30~g~cm^{-3}$.

As noted by Burrows \etal (2000), the fact that
HD~209458~b has a radius-to-mass ratio sufficiently higher than that of
Jupiter confirms the slowing of the radial shrinking process due to stellar
insolation.  They also point out that such a large radius requires the
planet to have migrated inward at a fairly early age if it were formed at
a distance beyond 0.5AU. A better handle on the planetary radius should
allow distinguishing between theoretical evolutionary models for
the planet itself, given different overall albedos.
With the discovery of more transiting planets in the near future and
with the $Kepler$ and $COROT$ missions, our ability to derive
good planetary parameters (radii, masses, \& densities) will depend
even more critically on the theoretical stellar models. The stellar
M/R relation is robust along most of the main sequence, and should provide
the best constraint for transit light curve solutions and $R_p$ determination.

\acknowledgements{ We are grateful to D. Guenther for providing special
runs with YREC for code comparisons.
We thank D. Charbonneau for making available to us the $HST$ transit light
curve before publication. We are grateful to D. Latham, R. Noyes and 
D. Charbonneau for reading the manuscript and many helpful discussions.
Thanks to the referee, P. Eggleton, for the good suggestions.
AMC wishes to thank the Harvard College Research Program for their continuing
support of her work; DDS acknowledges support from the Alfred P. Sloan
Foundation.}

\newpage
\singlespace
\begin{center}
\begin{deluxetable}{rccl}
\tablecaption{Evolutionary Model Data}
\tablewidth{0pc}
\tablecolumns{4}
\tablehead{
\colhead{\phm{$flahghghghgm$}$\log T_{\rm eff}$ (K)} & 
\colhead{$\log L/L_{\odot}$} & 
\colhead{$R/R_{\odot}$} & 
\colhead{$\log \rm Age$ (yrs.)\phm{$afdafj$} }}

\tablecomments{For each track of particular mass, metallicity, and
mixing-length parameter, five points on the evolutionary track are listed.
The first point denotes the zero-age main sequence, and the last marks the
end of the modeling sequence, while the three others correspond to the
minimum, central, and maximum luminosities of our error box ($\log
L/L_{\odot}=0.164$, $\log L/L_{\odot}=0.204$, $\log L/L_{\odot}=0.244$,
respectively).  In some cases, fewer than five points are given for tracks
starting out with high luminosity.}

\startdata
\cutinhead{$M=1.30M_{\odot}$, [Fe/H]=0.25, $X=0.70$, $Y=0.28$,
$Z=0.035$, $\alpha=1.69$}
3.7830 & 0.2288 & 1.1797 & 0.0000\\
3.7831 & 0.2443 & 1.2004 & 8.2425\\
3.7837 & 0.3398 & 1.3362 & 9.3603\\
\cutinhead{$M=1.25M_{\odot}$, [Fe/H]=0.25, $X=0.70$, $Y=0.28$,
$Z=0.035$, $\alpha=1.69$}
3.7738 & 0.1449 & 1.3985 & 0.0000\\
3.7742 & 0.1653 & 1.1419 & 8.4527\\
3.7761 & 0.2098 & 1.1914 & 9.1316\\
3.7768 & 0.2433 & 1.2343 & 9.3319\\
3.7763 & 0.2778 & 1.2873 & 9.4738\\
\cutinhead{$M=1.20M_{\odot}$, [Fe/H]=0.25, $X=0.70$, $Y=0.28$,
$Z=0.035$, $\alpha=1.69$}
3.7640 & 0.0575 & 1.0569 & 0.0000\\
3.7686 & 0.1614 & 1.1662 & 9.3975\\
3.7689 & 0.2070 & 1.2277 & 9.5663\\
3.7664 & 0.2400 & 1.2897 & 9.6682\\
3.7662 & 0.2388 & 1.2894 & 9.6693\\
\cutinhead{$M=1.25M_{\odot}$, [Fe/H]=0.18, $X=0.70$, $Y=0.28$,
$Z=0.030$, $\alpha=1.69$}
3.7839 & 0.1997 & 1.1361 & 0.0000\\
3.7840 & 0.2081 & 1.1466 & 7.5710\\
3.7852 & 0.2459 & 1.1910 & 8.9213\\
3.7852 & 0.3303 & 1.3126 & 9.4276\\
\cutinhead{$M=1.20M_{\odot}$, [Fe/H]=0.18, $X=0.70$, $Y=0.28$,
$Z=0.030$, $\alpha=1.69$}
3.7743 & 0.1125 & 1.0740 & 0.0000\\
3.7761 & 0.1603 & 1.1254 & 8.9752\\
3.7780 & 0.2095 & 1.1806 & 9.3240\\
3.7783 & 0.2454 & 1.2287 & 9.4699\\
3.7780 & 0.2620 & 1.2542 & 9.5230\tablebreak
\cutinhead{$M=1.18M_{\odot}$, [Fe/H]=0.18, $X=0.70$, $Y=0.28$,
$Z=0.030$, $\alpha=1.69$}
3.7703 & 0.0766 & 0.9611 & 0.0000\\
3.7741 & 0.1629 & 0.9444 & 9.2869\\
3.7752 & 0.2075 & 0.9396 & 9.4785\\
3.7745 & 0.2442 & 0.9427 & 9.5931\\
3.7738 & 0.2522 & 0.9457 & 9.6175\\
\cutinhead{$M=1.175M_{\odot}$, [Fe/H]=0.10, $X=0.70$, $Y=0.28$,
$Z=0.025$, $\alpha=1.69$}
3.7809 & 0.1295 & 0.9155 & 0.0000\\
3.7820 & 0.1661 & 0.9108 & 8.8061\\
3.7839 & 0.2104 & 0.9026 & 9.2142\\
3.7848 & 0.2450 & 0.8989 & 9.3785\\
3.7846 & 0.2766 & 0.8999 & 9.4886\\
\cutinhead{$M=1.15M_{\odot}$, [Fe/H]=0.10, $X=0.70$, $Y=0.28$,
$Z=0.025$, $\alpha=1.69$}
3.7758 & 0.0837 & 1.0319 & 0.0000\\
3.7793 & 0.1659 & 1.1161 & 9.2446\\
3.7807 & 0.2088 & 1.1656 & 9.4358\\
3.7806 & 0.2453 & 1.2157 & 9.5502\\
3.7780 & 0.2756 & 1.2740 & 9.6401\\
\cutinhead{$M=1.14M_{\odot}$, [Fe/H]=0.10, $X=0.70$, $Y=0.28$,
$Z=0.025$, $\alpha=1.69$}
3.7737 & 0.0652 & 1.0199 & 0.0000\\
3.7781 & 0.1628 & 1.1183 & 9.3341\\
3.7793 & 0.2100 & 1.1743 & 9.5097\\
3.7788 & 0.2440 & 1.2240 & 9.6051\\
3.7651 & 0.4366 & 1.6273 & 9.8359\\
\cutinhead{$M=1.13M_{\odot}$, [Fe/H]=0.10, $X=0.70$, $Y=0.28$,
$Z=0.025$, $\alpha=1.69$}
3.7716 & 0.0465 & 1.0079 & 0.0000\\
3.7770 & 0.1663 & 1.1285 & 9.4362\\
3.7777 & 0.2100 & 1.1829 & 9.5720\\
3.7765 & 0.2439 & 1.2368 & 9.6590\\
3.7671 & 0.4040 & 1.5530 & 9.8382\tablebreak
\cutinhead{$M=1.12M_{\odot}$, [Fe/H]=0.10, $X=0.70$, $Y=0.28$,
$Z=0.025$, $\alpha=1.69$}
3.7694 & 0.0276 & 0.9649 & 0.0000\\
3.7755 & 0.1613 & 0.9381 & 9.4944\\
3.7760 & 0.2081 & 0.9363 & 9.6241\\
3.7746 & 0.2428 & 0.9422 & 9.7024\\
3.7737 & 0.2780 & 0.9462 & 9.7524\\
\cutinhead{$M=1.10M_{\odot}$, [Fe/H]=0.10, $X=0.70$, $Y=0.28$,
$Z=0.025$, $\alpha=1.69$}
3.7651 & -0.0105 & 0.9726 & 0.0000\\
3.7727 & 0.1619 & 1.1453 & 9.6248\\
3.7720 & 0.2071 & 1.2104 & 9.7256\\
3.7709 & 0.2453 & 1.2712 & 9.7825\\
3.7584 & 0.3918 & 1.5939 & 9.9055\\
\cutinhead{$M=1.15M_{\odot}$, [Fe/H]=0.00, $X=0.70$, $Y=0.28$,
$Z=0.020$, $\alpha=1.69$}
3.7891 & 0.1546 & 1.0532 & 0.0000\\
3.7892 & 0.1649 & 1.0653 & 7.6434\\
3.7908 & 0.2067 & 1.1095 & 8.9677\\
3.7923 & 0.2454 & 1.1519 & 9.2369\\
3.7926 & 0.3071 & 1.2350 & 9.4692\\
\cutinhead{$M=1.10M_{\odot}$, [Fe/H]=0.00, $X=0.70$, $Y=0.28$,
$Z=0.020$, $\alpha=1.69$}
3.7788 & 0.0606 & 0.9911 & 0.0000\\
3.7834 & 0.1617 & 1.0899 & 9.3356\\
3.7848 & 0.2109 & 1.1460 & 9.5080\\
3.7848 & 0.2452 & 1.1923 & 9.5954\\
3.7767 & 0.4072 & 1.4910 & 9.8025\\
\cutinhead{$M=1.09M_{\odot}$, [Fe/H]=0.00, $X=0.70$, $Y=0.28$,
$Z=0.020$, $\alpha=1.69$}
3.7766 & 0.0413 & 0.9791 & 0.0000\\
3.7824 & 0.1653 & 1.0996 & 9.4361\\
3.7833 & 0.2054 & 1.1467 & 9.5546\\
3.7827 & 0.2459 & 1.2233 & 9.6494\\
3.7794 & 0.3397 & 1.3638 & 9.7757\tablebreak
\cutinhead{$M=1.06M_{\odot}$, [Fe/H]=0.00, $X=0.70$, $Y=0.28$,
$Z=0.020$, $\alpha=1.69$}
3.7699 & -0.0174 & 0.9438 & 0.0000\\
3.7782 & 0.1624 & 1.1173 & 9.6277\\
3.7780 & 0.2063 & 1.1763 & 9.7167\\
3.7769 & 0.2422 & 1.2322 & 9.7726\\
3.7622 & 0.4097 & 1.5989 & 9.9064\\
\cutinhead{$M=1.05M_{\odot}$, [Fe/H]=0.00, $X=0.70$, $Y=0.28$,
$Z=0.020$, $\alpha=1.69$}
3.7676 & -0.0372 & 0.9323 & 0.0000\\
3.7767 & 0.1661 & 1.1298 & 9.6891\\
3.7758 & 0.2085 & 1.1913 & 9.7662\\
3.7749 & 0.2434 & 1.2453 & 9.8094\\
3.7739 & 0.2736 & 1.2953 & 9.8388\\
\cutinhead{$M=1.04M_{\odot}$, [Fe/H]=0.00, $X=0.70$, $Y=0.28$,
$Z=0.020$, $\alpha=1.69$}
3.7653 &-0.0572 & 0.9208 & 0.0000\\
3.7749 & 0.1617 & 1.1335 & 9.7308\\
3.7739 & 0.2098 & 1.2036 & 9.8072\\
3.7731 & 0.2423 & 1.2541 & 9.8419\\
3.7716 & 0.2782 & 1.3160 & 9.8732\\
\cutinhead{$M=1.00M_{\odot}$, [Fe/H]=0.00, $X=0.70$, $Y=0.28$,
$Z=0.020$, $\alpha=1.69$}
3.7556 & -0.1375 &0.8778 & 0.0000\\
3.7673 & 0.1635 & 1.1760 & 9.8997\\
3.7661 & 0.2088 & 1.2460 & 9.9389\\
3.7642 & 0.2435 & 1.3085 & 9.9632\\
3.7587 & 0.2971 & 1.4274 & 9.9946\\
\cutinhead{$M=1.015M_{\odot}$, [Fe/H]=-0.09, $X=0.70$, $Y=0.28$,
$Z=0.016$, $\alpha=1.69$}
3.7725 & 0.9063 & 0.9063 & 0.0000\\
3.7819 & 0.1655 & 1.1023 & 9.6838\\
3.7816 & 0.2079 & 1.1589 & 9.7557\\
3.7806 & 0.2457 & 1.2163 & 9.8035\\
3.7713 & 0.3814 & 1.4843 & 9.9038\tablebreak
\cutinhead{$M=0.997M_{\odot}$, [Fe/H]=-0.09, $X=0.70$, $Y=0.28$,
$Z=0.016$, $\alpha=1.69$}
3.7682 & -0.0791 & 0.9705 & 0.0000\\
3.7790 & 0.1655 & 0.9233 & 9.7695\\
3.7782 & 0.2058 & 0.9269 & 9.8252\\
3.7773 & 0.2439 & 0.9307 & 9.8619\\
3.7683 & 0.3580 & 0.9702 & 9.9372\\
\cutinhead{$M=0.984M_{\odot}$, [Fe/H]=-0.09, $X=0.70$, $Y=0.28$,
$Z=0.016$, $\alpha=1.69$}
3.7650 & -0.1058 & 0.8718 & 0.0000\\
3.7766 & 0.1611 & 1.1237 & 9.8199\\
3.7757 & 0.2081 & 1.1914 & 9.8729\\
3.7746 & 0.2437 & 1.2475 & 9.9022\\
3.7725 & 0.2819 & 1.3162 & 9.9282\\
\cutinhead{$M=0.948M_{\odot}$, [Fe/H]=-0.20, $X=0.70$, $Y=0.28$,
$Z=0.0125$, $\alpha=1.69$}
3.7697 & -0.1160 & 0.9638 & 0.0000\\
3.7820 & 0.1652 & 0.9106 & 9.8273\\
3.7812 & 0.2101 & 0.9139 & 9.8749\\
3.7804 & 0.2429 & 0.9175 & 9.9005\\
3.7797 & 0.2594 & 0.9202 & 9.9119\\
\cutinhead{$M=0.921M_{\odot}$, [Fe/H]=-0.20, $X=0.70$, $Y=0.28$,
$Z=0.0125$, $\alpha=1.69$}
3.7628 & -0.1736 & 0.8145 & 0.0000\\
3.7768 & 0.1660 & 1.1290 & 9.9288\\
3.7758 & 0.2085 & 1.1913 & 9.9595\\
3.7744 & 0.2389 & 1.2419 & 9.9777\\
\cutinhead{$M=1.10M_{\odot}$, [Fe/H]=0.00, $X=0.72$, $Y=0.26$,
$Z=0.020$, $\alpha=1.69$}
3.7696 & 0.0060 & 0.9642 & 0.0000\\
3.7771 & 0.1612 & 0.9315 & 9.5681\\
3.7777 & 0.2083 & 0.9290 & 9.6760\\
3.7765 & 0.2452 & 0.9339 & 9.7467\\
3.7726 & 0.3382 & 0.9508 & 9.8459\tablebreak
\cutinhead{$M=1.06M_{\odot}$, [Fe/H]=0.00, $X=0.72$, $Y=0.26$,
$Z=0.020$, $\alpha=1.69$}
3.7606 & -0.0707 & 1.0052 & 0.0000\\
3.7710 & 0.1623 & 0.9578 & 9.7813\\
3.7699 & 0.2104 & 0.9628 & 9.8511\\
3.7689 & 0.2462 & 0.9674 & 9.8864\\
3.7673 & 0.2795 & 0.9746 & 9.9135\\
\cutinhead{$M=1.06M_{\odot}$, [Fe/H]=0.00, $X=0.68$, $Y=0.30$,
$Z=0.020$, $\alpha=1.69$}
3.7793 & 0.0380 & 0.9223 & 0.0000\\
3.7849 & 0.1644 & 0.8987 & 9.4339\\
3.7856 & 0.2044 & 0.8955 & 9.5502\\
3.7845 & 0.2442 & 0.9001 & 9.6443\\
3.7800 & 0.3653 & 0.9191 & 9.7847\\
\cutinhead{$M=1.02M_{\odot}$, [Fe/H]=0.00, $X=0.68$, $Y=0.30$,
$Z=0.020$, $\alpha=1.69$}
3.7701 & -0.0429 & 0.9620 & 0.0000\\
3.7789 & 0.1626 & 0.9237 & 9.6812\\
3.7779 & 0.2091 & 0.9281 & 9.7617\\
3.7772 & 0.2424 & 0.9311 & 9.7999\\
3.7735 & 0.3179 & 0.9470 & 9.8629\\
\cutinhead{$M=1.13M_{\odot}$, [Fe/H]=0.00, $X=0.70$, $Y=0.28$,
$Z=0.020$, $\alpha=1.40$}
3.7753 & 0.1165 & 1.0740 & 0.0000\\
3.7768 & 0.1621 & 1.1244 & 8.9145\\
3.7787 & 0.2094 & 1.1769 & 9.2654\\
3.7795 & 0.2454 & 1.2221 & 9.4114\\
3.7757 & 0.3133 & 1.3444 & 9.6076\\
\cutinhead{$M=1.06M_{\odot}$, [Fe/H]=0.00, $X=0.70$, $Y=0.28$,
$Z=0.020$, $\alpha=1.40$}
3.7600 & -0.0197 & 0.9852 & 0.0000\\
3.7671 & 0.1611 & 1.1739 & 9.6296\\
3.7662 & 0.2104 & 1.2476 & 9.7287\\
3.7648 & 0.2459 & 1.3085 & 9.7803\\
3.7595 & 0.3341 & 1.4839 & 9.8609\tablebreak
\cutinhead{$M=1.06M_{\odot}$, [Fe/H]=0.00, $X=0.70$, $Y=0.28$,
$Z=0.020$, $\alpha=2.00$}
3.7782 & -0.0148 & 0.9110 & 0.0000\\
3.7876 & 0.1649 & 1.0730 & 9.6270\\
3.7877 & 0.2081 & 1.1272 & 9.7149\\
3.7869 & 0.2438 & 1.1787 & 9.7703\\
3.7837 & 0.3374 & 1.3324 & 9.8568\\
\cutinhead{$M=1.00M_{\odot}$, [Fe/H]=0.00, $X=0.70$, $Y=0.28$,
$Z=0.020$, $\alpha=2.00$}
 3.7648 & -0.1239 & 0.9857 & 0.0000\\
3.7786 & 0.1636 & 0.9252 & 9.8762\\
3.7780 & 0.2107 & 0.9278 & 9.9198\\
3.7768 & 0.2444 & 0.9326 & 9.9445\\
3.7760 & 0.2596 & 0.9361 & 9.9545\\
\cutinhead{$M=1.089M_{\odot}$, [Fe/H]=0.00, $X=0.70$, $Y=0.28$,
$Z=0.020$, $\alpha=1.55$}
3.7719 & 0.0385 & 0.9540 & 0.0000\\
3.7773 & 0.1631 & 0.9304 & 9.4391\\
3.7782 & 0.2093 & 0.9267 & 9.5725\\
3.7775 & 0.2440 & 0.9299 & 9.6523\\
3.7714 & 0.3690 & 0.9563 & 9.8047\\
\cutinhead{$M=1.06M_{\odot}$, [Fe/H]=0.00, $X=0.70$, $Y=0.28$,
$Z=0.020$, $\alpha=1.55$}
3.7654 & -0.0186 & 0.9828 & 0.0000\\
3.7731 & 0.1617 & 0.9486 & 9.6286\\
3.7728 & 0.2059 & 0.9502 & 9.7180\\
3.7714 & 0.2466 & 0.9564 & 9.7793\\
3.7667 & 0.3351 & 0.9770 & 9.8599\\
\cutinhead{$M=1.06M_{\odot}$, [Fe/H]=0.00, $X=0.70$, $Y=0.28$,
$Z=0.020$, $\alpha=1.85$}
3.7745 & -0.0160 & 0.9255 & 0.0000\\
3.7833 & 0.1639 & 1.0934 & 9.6279\\
3.7833 & 0.2073 & 1.1494 & 9.7159\\
3.7822 & 0.2430 & 1.2034 & 9.7714\\
3.7787 & 0.3364 & 1.3618 & 9.8578\tablebreak
\cutinhead{$M=1.033M_{\odot}$, [Fe/H]=0.00, $X=0.70$, $Y=0.28$,
$Z=0.020$, $\alpha=1.85$}
3.7681 & -0.0696 & 0.8962 & 0.0000\\
3.7789 & 0.1626 & 1.1138 & 9.7623\\
3.7782 & 0.2090 & 1.1788 & 9.8291\\
3.7774 & 0.2453 & 1.2334 & 9.8647\\
3.7748 & 0.3000 & 1.3297 & 9.9067\\

\enddata
\end{deluxetable}

\newpage

\begin{figure} 
\plotone{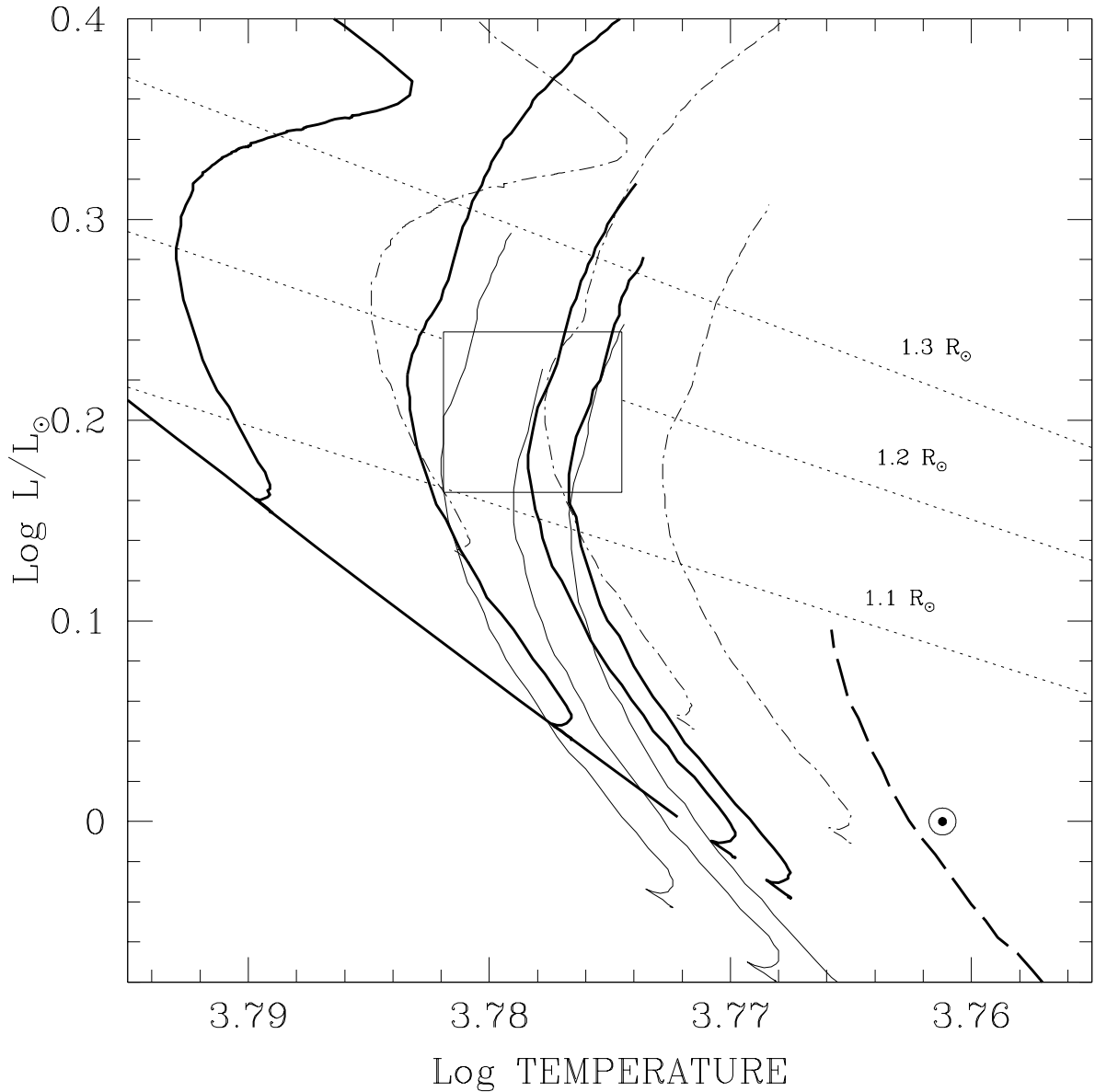}

\caption{A small section of the Hertzsprung-Russell diagram in the
vicinity of HD~209458 and the Sun. Our computed evolution tracks are
shown for $Z$~=~0.02 ({\em thick solid lines}), for $Z$~=~0.016
({\em thin solid lines}), and for $Z$~=~0.025 ({\em dot-dash lines}).
Masses range from 1.05 to 1.15$M_{\odot}$ for $Z$~=~0.02 (see text).
The long-dashed track is the theoretical evolution of the Sun,
while the dot near it marks the known solar temperature and luminosity
values. All tracks that do not run off the plot are stopped at 7 Gyr.
Three lines of constant radius are shown as well.}

\end{figure}

\newpage

\begin{figure}[!t]
\plotone{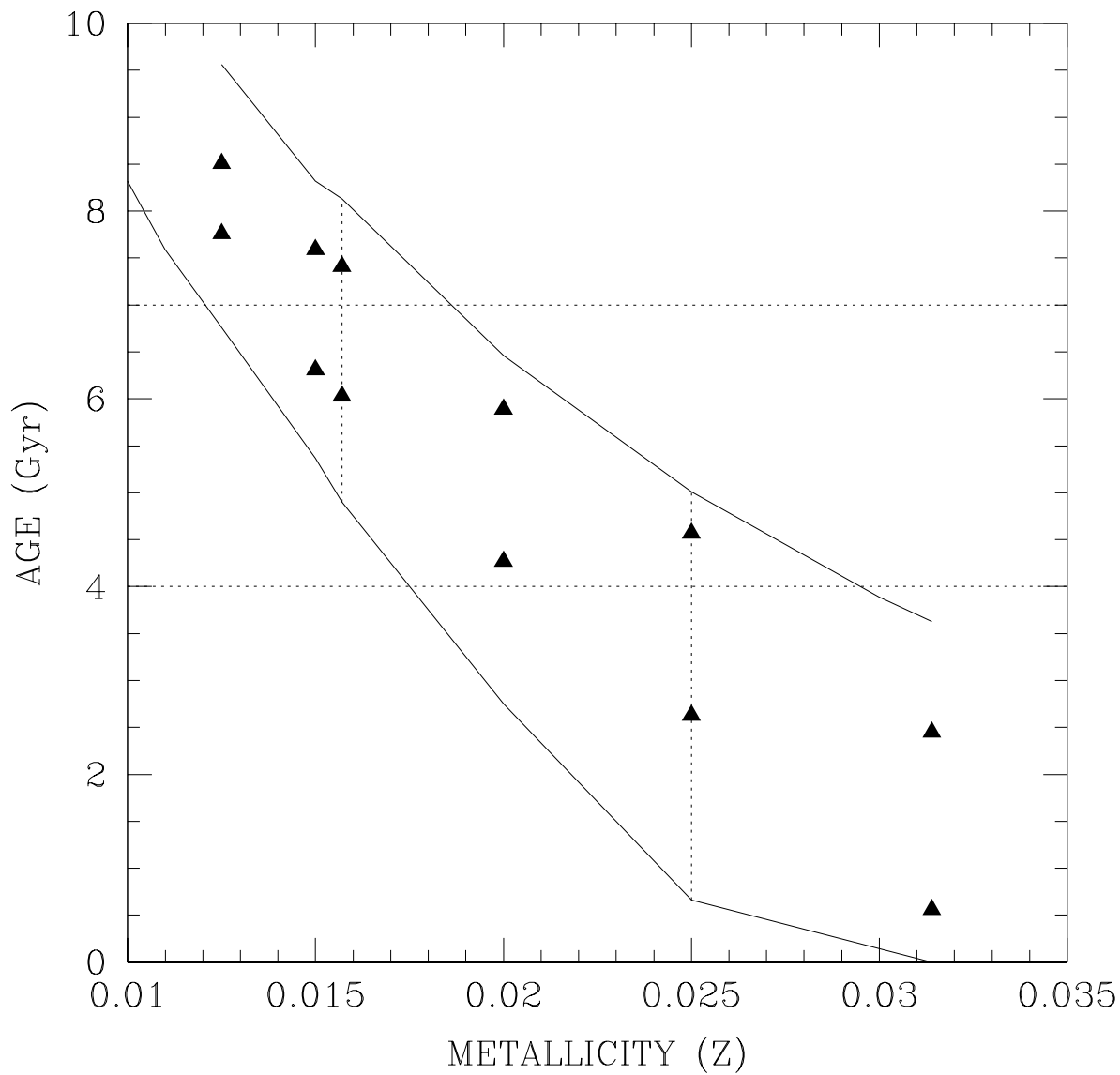}

\caption{The correlation between stellar metallicity and computed age. For
selected metallicities, two triangles denote the ages at which the
"best" stellar model (the one that achieves the target values $T_{\rm
eff}=6000~K$  and $L/L_{\odot}=1.61$) evolves through the limits of
the temperature/luminosity error range for HD~209458. The curves illustrate 
maximum and minimum possible ages over all models, and horizontal lines
mark our preferred age bounds used in Figs. 3 \& 4.}

\end{figure}

\newpage 
\begin{figure}[!t] 
\plotone{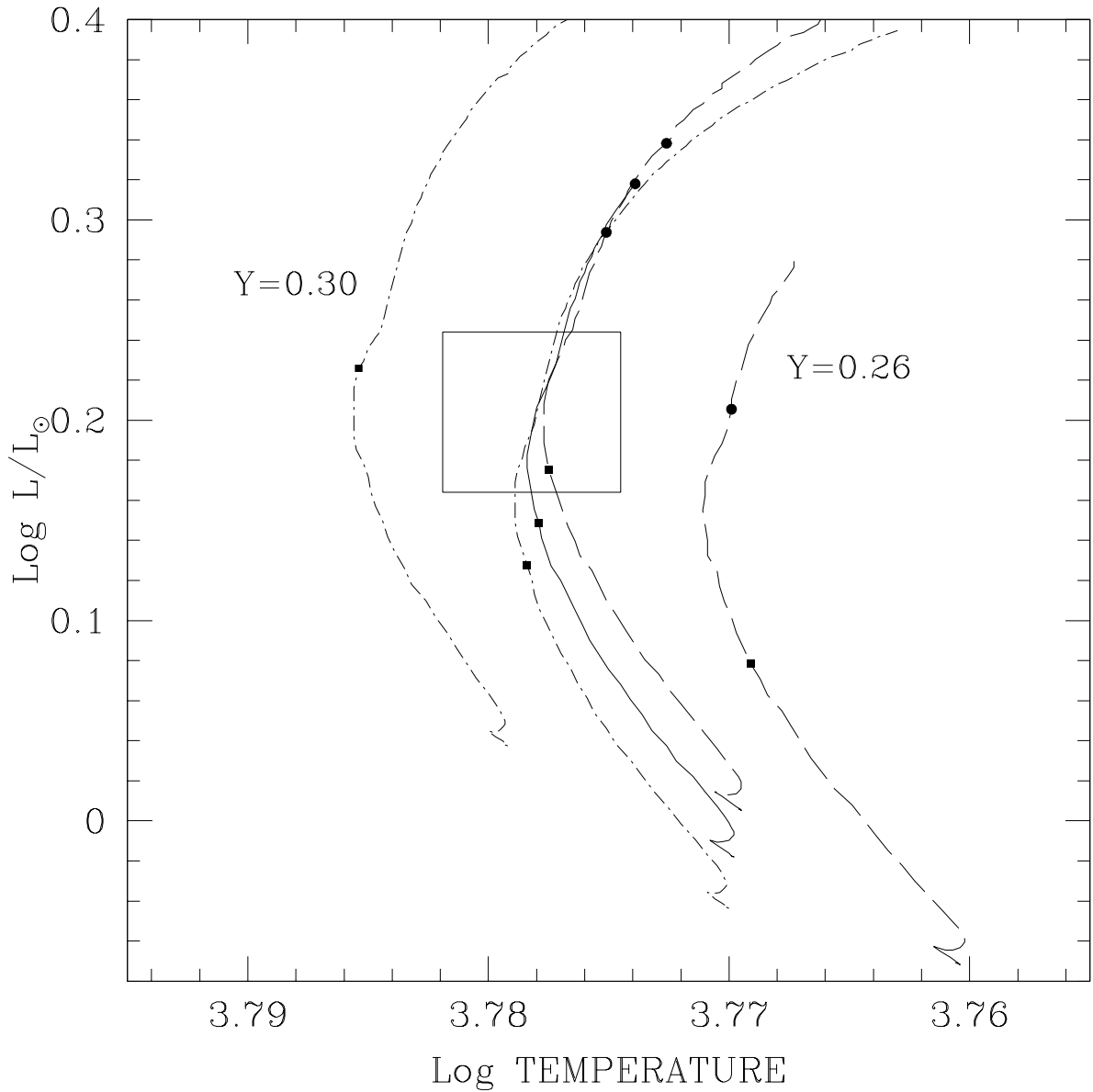}

\caption{The effect of changing the initial helium abundance on the models
in the same region of the H-R diagram shown in Fig. 1.
The patterns trace tracks of fixed $Y$-value
(as marked), and the central, unbroken curve corresponds to the most
favored model of $Y$~=~0.28, $M=1.06M_{\odot}$. The middle $Y$~=~0.26
model has mass $1.10~M_{\odot}$, while the middle $Y$~=~0.30 model has
mass $1.02~M_{\odot}$; the two on the outside are both
$1.06~M_{\odot}$.  All models are for a metallicity $Z$~=~0.02. In
addition, ages are delimited by squares and circles, which indicate
the 4-Gyr and the 7-Gyr points, respectively.}

\end{figure}

\newpage
\begin{figure}[!t]
\plotone{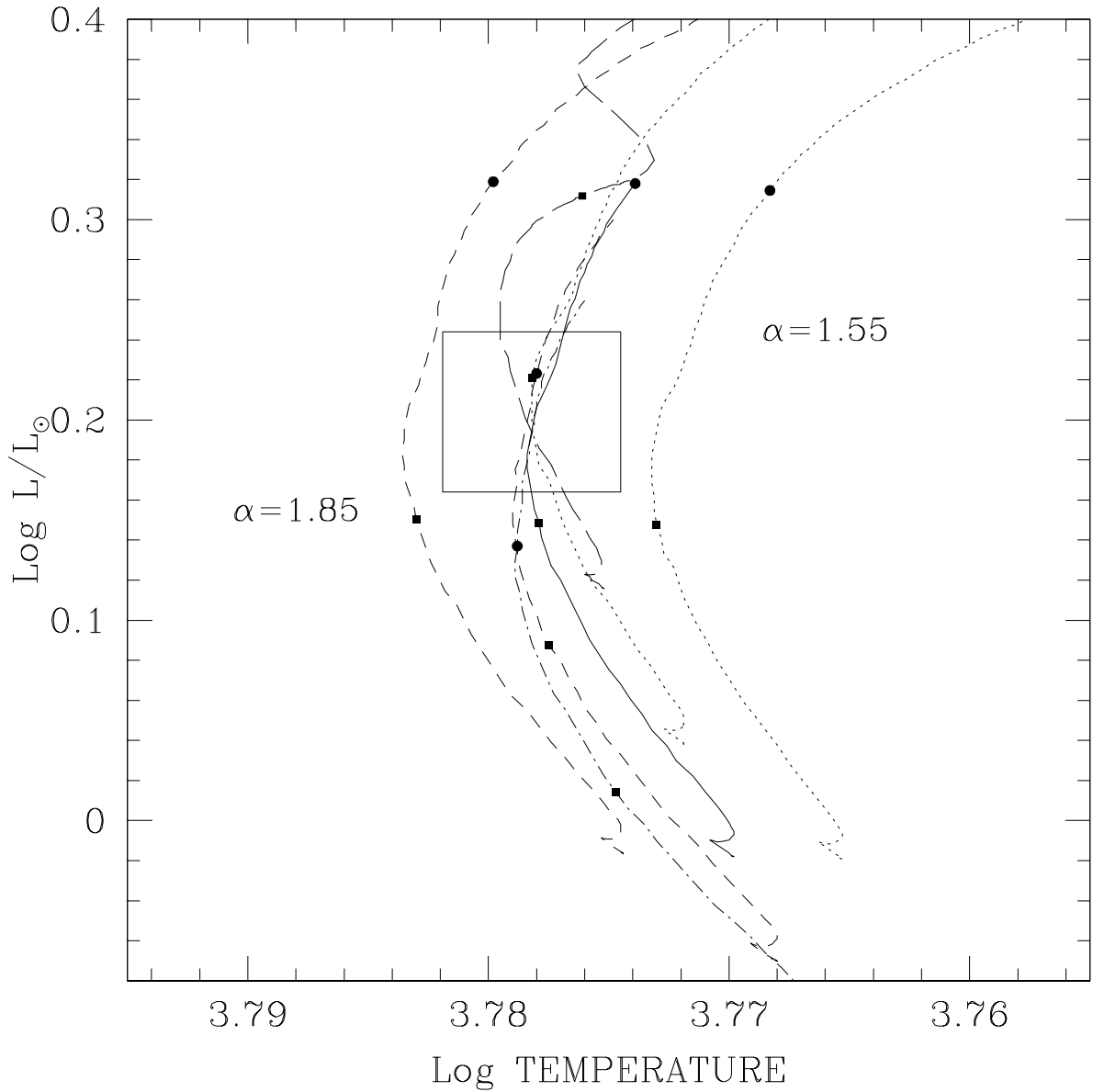}

\caption{The effect of changing the mixing length parameter on the models
in the same region of the H-R diagram shown in Fig. 1.  All models
plotted here have abundances $Z$~=~0.02 and $Y$~=~0.28.  The dotted and
short-dashed curves have $\alpha$-values as marked, the dot-dash pattern
traces a model of $\alpha=2.00$, $M=1.01M_{\odot}$, and the
long-dashed track indicates a model with $\alpha=1.40$,
$M=1.13M_{\odot}$. The solid curve is a track maintaining the solar
$\alpha$ (1.69) and $1.06~M_{\odot}$, while the inner $\alpha=1.55$ and
$\alpha=1.85$ tracks have had their masses shifted to $1.09~M_{\odot}$
and $1.03~M_{\odot}$, respectively.  The two outside tracks shown are
$1.06~M_{\odot}$.  As in figure 3, squares mark the
4~Gyr point, and circles denote 7~Gyr.}

\end{figure}

\newpage
\begin{figure}[!t]
\plotone{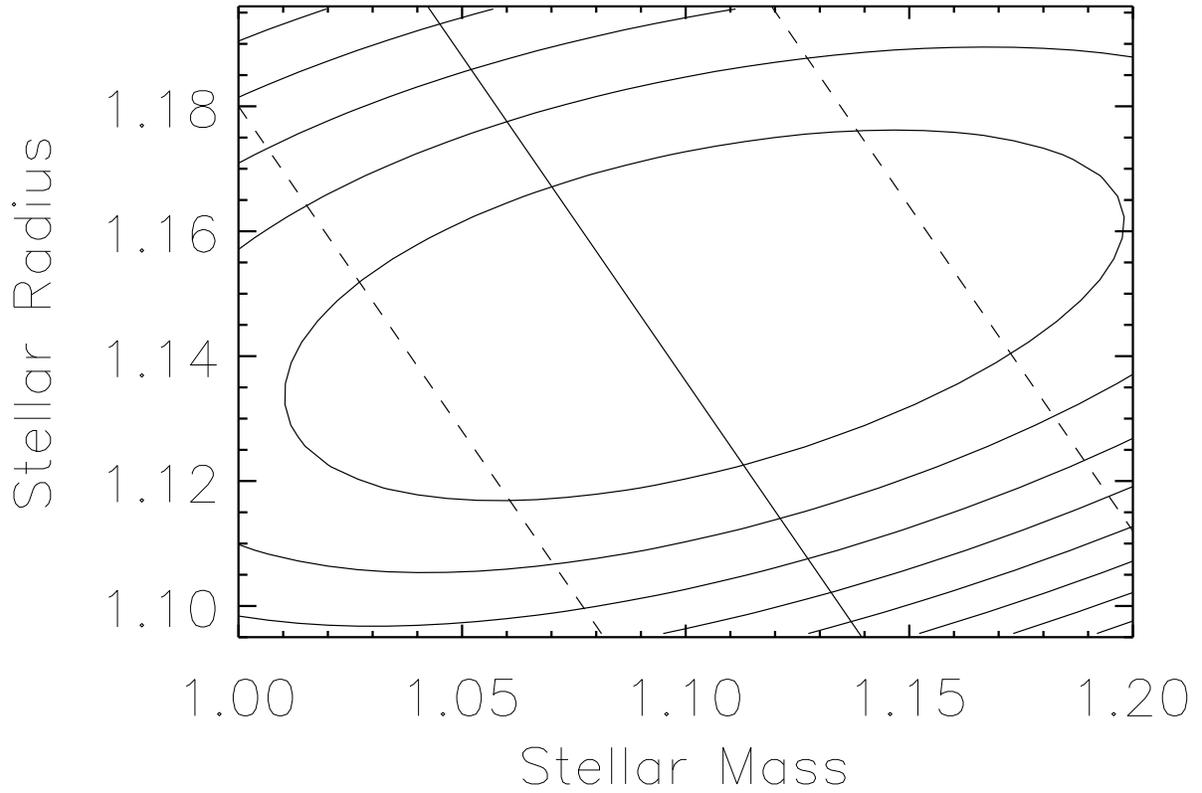}

\caption{The astrophysical relation between stellar mass and radius
(solid straight line) breaks the degeneracy in the transit light curve
solution, which produces an almost orthogonal relation (curved contours)
taken from Brown et al. (2001). All units are solar.
}
\end{figure}

\end{center}

\end{document}